\documentclass[aps,prb,twocolumn,showpacs,a4paper,unsortedaddress,
amsmath,amssymb,byrevtex]{revtex4}

\usepackage[latin1]{inputenc}
\usepackage{graphicx}
\usepackage{dcolumn}
\usepackage{bm}
\usepackage{hyperref}
\usepackage{times}

\begin{document}
\preprint{ffuov/02-01}

\title{Electronic Properties of Alkali- and Alkaline-Earth-Intercalated
Silicon Nanowires}

\author{S. Sirichantaropass}
\author{V. M. Garc\'{\i}a-Su\'arez}\email{v.garcia-suarez@lancaster.ac.uk}
\author{C. J. Lambert}
\affiliation{Department of Physics, Lancaster University,
Lancaster, LA1 4YB, U. K.}

\date{\today}

\begin{abstract}
We present a first-principles study of the electronic properties
of silicon clathrate nanowires intercalated with various types of
alkali or alkaline-earth atoms. We find that the band structure of
the nanowires can be tailored by varying the impurity atom within
the nanowire. The electronic character of the resulting systems
can vary from metallic to semiconducting with direct band gaps.
These properties make the nanowires specially suitable for
electrical and optoelectronic applications.
\end{abstract}

\pacs{73.22.-f,73.21.Hb,81.07.Vb}

\maketitle

The desire for nanoscale components which integrate gracefully
with silicon CMOS technology makes the fabrication and
characterization of silicon nanowires particularly attractive.
These structures can be grown using a broad range of experimental
techniques \cite{Zha98,Mor98,Mar99,Gol00,Hol00,Ma03} and posses
novel properties which may make them suitable as interconnects in
very-large-scale integrated devices. Other interesting
applications include photoelectronics, since these structures have
large direct band-gaps which allow them to work as visible-light
emitters with low power consumption. The origin of this change in
the band gap is related to quantum confinement \cite{Wan00,Zha04}
which increases the band width and produces an indirect-direct
transition as the size of the wire shrinks \cite{Pon06}. The band
gap generates a low-voltage gap in the I-V characteristic which is
considerably enhanced when the surface is passivated \cite{Dur06}.
However, upon doping the gap disappears and the resulting I-V
curves display ohmic behavior for low voltages \cite{Zhe04}.

\begin{figure}
\includegraphics[width=\columnwidth]{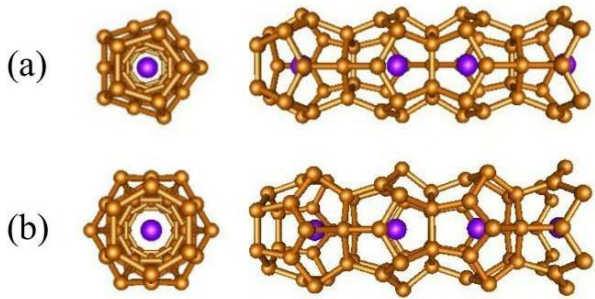}
\caption{\label{Fig1}Top and lateral views of two unit cells of a
2M@Si(30) (a) and a 2M@Si(36) (b), where M is an alkali or
alkaline-earth atom.}
\end{figure}

Many possible structures for silicon nanowires have been proposed
\cite{Men99,Mar99,Li02,Zha03,Zha04,Men04,Rur05,Kag05,Pon05,Dur06,Pon06}.
Those grown from porous silicon are the most stable for diameters
down to 1 nm \cite{Kag05,Pon05,Pon06}. Their structural and
cohesive properties are determined by a competition between the
highly coordinated core and the surface reconstruction which, due
to curvature effects, becomes more important as the nanowire width
decreases. There are, however, other types of silicon nanowires
with a hollow core and a more coordinated surface which resemble
fullerene-like structures. Some of such nanowires can be made
stable by the encapsulation of metallic particles
\cite{And02,Sin03,Kum04}, which alter significantly their physical
properties and transform the nanowires from semiconducting to
metallic. However, pure silicon nanowires derived from the silicon
clathrate phases do not need any impurity to be stabilized and are
expected to be the most stable for very small diameters
\cite{Mar99,Kag05}. These nanowires can have many different shapes
\cite{Kag05,Pon05,Dur06,Pon06} depending on the basic unit of
repetition and the growth direction.

Clathrates are particularly interesting due to their novel elastic
\cite{Mig99}, thermoelectric \cite{Tse00}, optoelectronic
\cite{Gry00} and superconducting \cite{Kaw95} properties. They are
grown by nucleating a vapor of silicon or other group IV elements
like Ge and Sn around alkali or alkaline-earth atoms generally
from the third or bigger rows \cite{Mig05}. The impurities can be
left inside the structure with different concentrations or be
removed afterwards. Pristine clathrates are semiconducting and
have wider band gaps than the diamond phase of silicon
\cite{Ada94}, whereas intercalated clathrates are usually metallic
\cite{Mor00}. The most common clathrate lattices have 34 and 46
atoms in the primitive unit cell and are made of Si$_{20}$ and
Si$_{28}$ cages and Si$_{20}$ and Si$_{24}$ cages, respectively.

The smallest of the clathrate-type nanowires are based on the
Si$_{20}$ and Si$_{24}$ cages \cite{Mar99,Dur06} and have been
predicted to be the most stable configurations in some experiments
\cite{Mar99}. These nanowires are expected to show many of the
clathrate properties due to their similarities with the bulk
phases. The Si$_{20}$ cage, which is present in both types of
clathrates, is a regular polyhedron made of 12 pentagons. The
corresponding nanowire is grown along an axis that passes through
two opposite faces ($C_{5v}$ symmetry) and has 30 atoms in the
unit cell. The Si$_{24}$ cage has 12 pentagons and two opposite
hexagons. The corresponding nanowire is grown along the direction
perpendicular to the hexagons ($C_{6v}$ symmetry) and contains 36
atoms in the unit cell. The transport properties of this nanowire
placed between aluminium electrodes and passivated with hydrogen
were calculated by Landman and coworkers \cite{Lan00} and they
found that the nanowire doped with Al atoms was more efficient
than the pristine one \cite{FiniteWires}. We label these nanowires
as Si(30) and Si(36) respectively. The intercalation of alkali or
alkaline-earth impurities is made possible by the large endohedral
space available inside the cages and the fact that the
corresponding bulk structures are stable.

In this article we report ab-initio theoretical studies of the
electronic properties of Si(30) and Si(36) nanowires intercalated
with the alkali and alkaline-earth metals $\{$Na,K,Rb,Cs$\}$ and
$\{$Ca,Sr,Ba$\}$, respectively, which correspond to typical
clathrate intercalations \cite{Mig05}. We label them as 2M@Si(30)
and 2M@Si(36), where 2M represents the two alkali or alkaline
earth atoms included in the unit cell. An example of the top and
lateral views of these structures can be seen in Fig.
(\ref{Fig1}). We predict that intercalation alters dramatically
the electronic structure of the nanowires and allows one to tailor
their electronic properties. The overall changes can be traced
back to three different effects: structural deformation,
hybridization and charge transfer from the endohedral impurity. As
a consequence, the electronic character of the nanowires can range
from metallic to semiconducting. The band gaps in the latter are
usually direct and comparable to the band gap of bulk silicon,
which make these nanowires particularly useful for optoelectronics
applications in the infrared or visible range.

\begin{figure}
\includegraphics[width=\columnwidth]{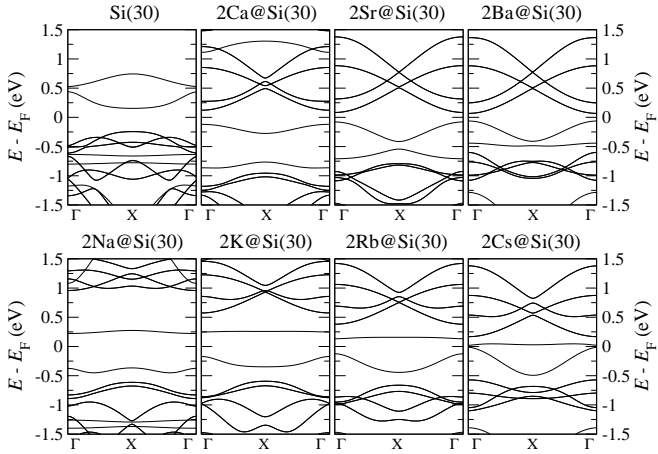}
\caption{\label{Fig2}Band structure of pristine (first top box),
alkaline-earth intercalated (next three top boxes) and alkali
intercalated (bottom) Si(30) nanowires.}
\end{figure}

The electronic structure was calculated with the SIESTA code
\cite{Sol02}, which is based on density functional theory
\cite{Koh65} and employs norm-conserving pseudopotentials and
linear combinations of atomic orbitals. The valence states were
spanned with optimized double-$\zeta$ polarized basis sets. In the
heaviest atoms (K,Ca,Rb,Sr,Cs,Ba) we included relativistic
corrections in the pseudopotentials and semicore states in the
basis set. The exchange and correlation energy and potential were
evaluated with the local density approximation (LDA). All atomic
coordinates and the lattice were fully relaxed until all the
forces and the stress were smaller than 0.02 eV/\AA\/ and 1.0 GPa,
respectively. The dimensions of the unit cell along the
perpendicular directions were made large enough to avoid overlaps
and electrostatic interactions with other images. The real space
grid was defined with a plane wave cutoff of 200 Ry. The Brillouin
zone along the nanowire growth direction was sampled with 30
$k$-points to perform the structural relaxations and 200
$k$-points to calculate the densities of states (DOS), where a
broadening parameter of 0.01 eV was used. The binding energies
were computed by removing the basis set superposition error
\cite{Boy70} which comes from the localized character of the basis
functions.

\begin{figure}
\includegraphics[width=\columnwidth]{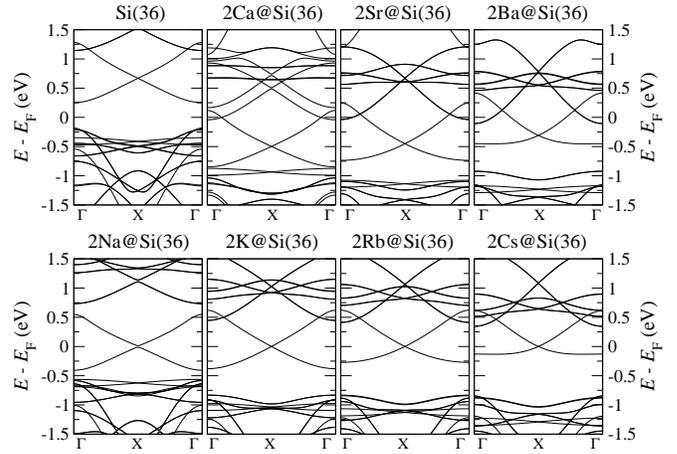}
\caption{\label{Fig3}Band structure of pristine (first top box),
alkaline-earth intercalated (next three top boxes) and alkali
intercalated (bottom) Si(36) nanowires.}
\end{figure}

First we focus on the pristine nanowires, whose initial unrelaxed
configuration corresponds to that of the ideal polyhedric
structures. Upon full relaxation the structure changes slightly
and compresses, resulting in lattice constants along the growth
direction of 10.50 \AA\/ and 10.28 \AA\/ for the Si(30) and
Si(36), respectively. The cohesive energies of both nanowires are
very similar, as can be seen in Table \ref{Tab01}, but the Si(36)
is slightly more stable \cite{Mar99}, which is probably due to the
fact that in this nanowire the presence of sixfold rings generates
bond angles much closer to the ideal bond angles of diamond
silicon and reduces the strain of the structure. The electronic
band structures computed for the pristine Si(30) and Si(36)
nanowires are shown in the first boxes of Figs. (\ref{Fig2}) and
(\ref{Fig3}), respectively. The Si(30) nanowire has a clear direct
band gap at the X point with a LDA value of 0.40 eV. Taking into
account that the LDA underestimates bands gaps, but gives correct
relative orders and trends, the previous result would imply, after
a scissor-correction of 0.7 eV, which is usually employed in
silicon \cite{Ada94}, a true band gap of $\sim$ 1.10 eV
\cite{GWSICLDAU}, similar to the band gap of diamond silicon. The
Si(36) nanowire has also a direct band gap, with a width of 0.44
eV \cite{LDAgap}, but it is located at $\Gamma$ instead of X. The
scissor correction gives in this case a value of 1.14 eV. The
character and width of such band gaps, which lie close to the low
visible energy range, make these nanowires perfect candidates for
light emitting devices in optoelectronic applications on the
nanoscale.

\begin{table}
\caption{Binding energies (BE) and band gaps (BG) of pristine and
alkali- or alkaline-earth intercalated Si(30) and Si(36)
nanowires.}\label{Tab01}
\begin{ruledtabular}
\begin{tabular}{lcccc}
\multicolumn{1}{l}{}&\multicolumn{2}{c}{Si(30)}
&\multicolumn{2}{c}{Si(36)}\\
&BE (eV)&BG (eV)&BE (eV)&BG(eV) \\
\hline
Pristine&-5.22&0.40&-5.25&0.44\\
\hline
Na&-2.25&0.60&-2.42&0.03\\
K&-1.40&0.42&-2.47&0.00\\
Rb&-0.72&0.26&-2.37&0.00\\
Cs&0.42&0.03&-2.08&0.00\\
\hline
Ca&-3.70&0.24&-4.88&0.00\\
Sr&-3.86&0.16&-4.52&0.00\\
Ba&-4.33&0.13&-5.67&0.00\\
\end{tabular}
\end{ruledtabular}
\end{table}

When the impurity atoms are inserted in the structure the lattice
undergoes a small deformation and expands. The changes in the mean
atomic distances and in the lattice vector along the growth
direction are in general smaller than a 5\%. In table \ref{Tab01}
we also show the magnitude of the band gaps and the binding energy
of the intercalated nanowires. All these structures are
exothermic, except for 2Cs@Si(30), which is slightly endothermic
\cite{BindingLDA}. Inspection of the binding energies shows three
clear trends. First, the Si(30) nanowires are always less stable
than the Si(36) due to the unfavorable interaction produced by the
strong compression in the small Si(30) cages. Second, the alkali
intercalations are always less stable than the alkaline-earth
because the radii of the alkali elements are larger than those of
the corresponding alkaline-earth and this again increases the
possibility of unfavorable interactions. Third, in the alkali
atoms bigger than sodium the binding energy decreases as the
atomic radius increases whereas in the alkaline-earth the trend is
just the opposite, with the exception of 2Sr@Si(36). These
apparently contradictory behaviors can be again understood in
terms of the bigger size of the alkali impurities as compared to
the alkaline-earth impurities and the fact that the latter
elements donate more charge to the silicon lattice, which
stabilizes the structure \cite{Tse00} and increases the bonding
with the positive ion.

The band structures of the intercalated nanowires are plotted in
Figs. (\ref{Fig2}) and (\ref{Fig3}). In the Si(30) structures, all
nanowires have a direct band gap at $\Gamma$. The only exception
to this rule is that corresponding to the lightest element,
2Na@Si(30), where the band gap is indirect, although it is
difficult to say due to the small difference between both maxima
of the top of the valence band. Moving to the bottom of the
periodic table, the band gap gradually decreases towards Cs and
Ba. This is due to the growing separation in energy between the
outer levels of the impurities and the silicon network levels
which decreases the interaction between them. It is interesting to
note that in the alkali atoms the magnitude of the gap is reduced
by the presence of a low-dispersive state located just above the
Fermi level, whereas in the alkaline-earth elements the lowest
conduction bands are very dispersive. Such features, like the
existence of gaps and low dispersive bands, would have important
consequences in the transport properties of these systems, giving
rise to interesting effects like negative differential resistance.
In the Si(36) structures the general situation is rather
different. Due to the larger size of the Si(36) cages and the
corresponding smaller interaction of the metallic atoms with the
silicon structure, the bands of the filled nanowires resemble more
closely those of the empty ones. For example, in 2K@Si(36),
2Rb@Si(36) and 2Cs@Si(36) the only effect seems to be a rigid
downwards shift of the whole silicon band structure. These
intercalations give rise to a metallic or semi-metallic behavior,
where two bands cross at the Fermi level or open a small band gap.
The same behavior is found for all the alkaline-earth
intercalations.

\begin{figure}
\includegraphics[width=\columnwidth]{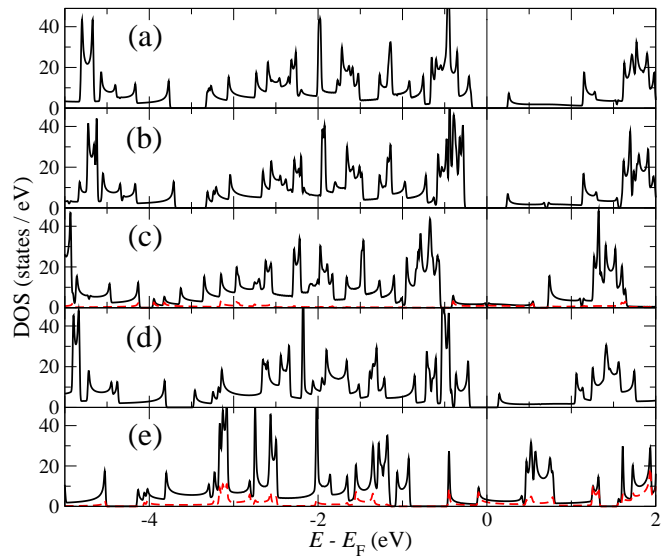}
\caption{\label{Fig4}Projected density of states on the silicon
atoms of a pristine Si(36) (a), a Na-deformed pristine Si(36) (b),
a 2Na@Si(36) (c), a Ba-deformed pristine Si(36) (d) and a
2Ba@Si(36) (e). The dashed lines in graphs (c) and (e) are the
projected densities of states on the sodium and barium atoms,
respectively, multiplied by a factor of 3 for clarity.}
\end{figure}

The above properties can be understood by dividing the effect of
the impurities into three parts: a structural deformation and two
purely electronic effects associated with charge transfer and
hybridization. The effect of the structural deformation can be
clearly determined by comparing the electronic structure of a
relaxed pristine clathrate with the electronic structure of a
pristine clathrate in the structural configuration of the
intercalated system. We show in Fig. (\ref{Fig4}) the density of
states of the pristine Si(36) nanowire and two typical
intercalations corresponding to the most studied intercalated
clathrates, i. e. those with sodium and barium \cite{Mor00}. Box
(a) corresponds to the pristine and relaxed nanowire, (b) and (d)
to the pristine nanowire in the structure of the Na- and
Ba-intercalated systems, respectively, and (c) and (e) to the
corresponding intercalated nanowires. As can be seen by comparing
graphs (a) with (b) or (d), structural modifications are more
pronounced in the barium encapsulation due to the larger size of
this element. In general, the bigger the atom, the larger the
modification it produces on the silicon network. The same behavior
is found for the Si(30) nanowires, but due to the smaller
endohedral space of the Si$_{20}$ cages and therefore the larger
deformation produced by the impurity, the changes are more
dramatic. These structural-induced changes in the electronic
properties can be understood in terms of the increase or, in some
cases, decrease of the distance between neighbor atoms, which
reduce or increase the couplings and therefore the bandwidths. The
final outcome of the structural deformations is then mainly a
modification of the size of the band gaps and widths. In some
cases, modification of the bond angles can also move states
downwards or upwards depending on whether or not they approach the
ideal angles of the $sp^3$ bonding. This is specially relevant in
the Si(30) nanowires.

The effect of the charge transfer can be easily demonstrated by
comparing the electronic band structures of the intercalated
systems with the pristine ones. Since alkali and alkaline-earth
atoms are very electropositive elements, their electronic levels
are well above those of silicon and tend to donate the outer one
(in the alkali) or two (in the alkaline-earth) electrons to the
silicon network. This can be clearly seen in Figs. (\ref{Fig2})
and (\ref{Fig3}), where one or two silicon bands move below the
Fermi level in the alkali or alkaline-earth intercalations,
respectively. Note that since the bands are not spin-split, the
number of electrons transferred per unit cell is 2 in the former
elements and 4 in the latter, coming from the two atoms in the
unit cell.

The hybridization between silicon states and impurity states can
be determined by examining the electronic density of states of
Fig. (\ref{Fig4}). As expected, large-diameter elements, such as
Ba, interact more strongly with the cage and produce greater
hybridization. This significantly modifies the electronic
structure of the intercalated system, compared to that of the
isostructural system without the impurity, at the energies where
the bands of the endohedral atom are located, as can be easily
deduced by looking at the projected density of states on the
barium atoms. As a consequence of the presence or absence of
hybridization for heavy or light elements, respectively, we
conclude that the bond between the bigger impurities and the
silicon network in the Si(36), and to a greater extent in the
Si(30), has some covalent character but becomes increasingly ionic
as the size of the atom decreases \cite{Charge}.

In summary, we have found that the combination of structural
deformation, charge transfer and hybridization in alkali or
alkaline-earth intercalated silicon clathrate nanowires produces a
rich behavior that allows one to tailor these systems for
technological applications. Their electronic character ranges from
semiconducting to metallic, which makes them suitable to act as
interconnects for nanocircuits or as other types of electronic
elements. Furthermore, the character and width of the band gaps,
which in most cases are direct and close to the visible, promise
important optoelectronic applications. Finally, the resemblance of
these systems to the clathrate bulk phases suggests that
superconductivity and novel thermoelectric or elastic properties
may be fruitful avenues for future investigations.

\begin{acknowledgments}
We acknowledge financial support from the European Commission and
the British EPSRC, DTI, Royal Society, and NWDA. VMGS thanks Jaime
Ferrer for useful discussions and the EU network
MRTN-CT-2004-504574 for a Marie Curie grant.
\end{acknowledgments}


\begin{thebibliography}{99}
\bibitem{GWSICLDAU} There are other methods that are known to
improve the energy gap and give quasiparticle energies close to
experiments, like the GW self-energy method \cite{Hyb86}, self
interaction corrections \cite{Per81}, and the LDA+U method
\cite{Ani97}. The application of some of these approaches to
silicon nanowires will be the topic of future investigations.
\bibitem{LDAgap} The LDA values are larger but comparable to those
found recently by Durandurdu \cite{Dur06} using GGA.
\bibitem{BindingLDA} Since the LDA tends to overestimate the biding
energies we do not exclude the possibility that some of the less
exothermic compounds, like 2K@Si(30) and 2Rb@Si(30), could be
slightly endothermic.
\bibitem{Charge} The hybridization and the prevalence of covalent or
ionic character can also be determined by looking at the charge
density around the metallic impurities \cite{Mor00}, i. e. if the
charge is smoothed out it means that the bond is stronger and has
a larger covalent character, whereas if the charge is strongly
localized around the impurity the bond is mainly ionic.
\end{thebibliography}
\end{document}